\documentstyle[psfig]{aipproc}

\def\la{\mathrel{\hbox{\rlap{\hbox{\lower4pt\hbox{$\sim$}}}\hbox{$<$}}}}
\def\ga{\mathrel{\hbox{\rlap{\hbox{\lower4pt\hbox{$\sim$}}}\hbox{$>$}}}}

\begin{document}
\title{Gamma-Ray Bursts as a Probe of the Very High Redshift Universe}

\author{Donald Q. Lamb and Daniel E. Reichart} 
\address{Department of Astronomy \& Astrophysics, University of Chicago,
\\ 5640 South Ellis Avenue, Chicago, IL 60637}

\maketitle

\begin{abstract}    
We show that, if many GRBs are indeed produced by the collapse of
massive stars, GRBs and their afterglows provide a powerful probe of
the very high redshift ($z \gtrsim 5$) universe.  
\end{abstract}

\section*{Introduction} 

There is increasingly strong evidence that gamma-ray bursts (GRBs) are
associated with star-forming galaxies [1,2,3,4] and occur near or in
the star-forming regions of these galaxies [5,3,4,6,2].  These
associations provide indirect evidence that at least the long GRBs
detected by BeppoSAX are a result of the collapse of massive stars. 
The discovery of what appear to be supernova components in the
afterglows of GRBs 970228 [7,8] and 980326 [9] provides direct evidence
that at least some GRBs are related to the deaths of massive stars, as
predicted by the widely-discussed collapsar model of GRBs
[10,11,12,13,14,15].  If GRBs are indeed related to the collapse of
massive stars, one expects the GRB rate to be approximately
proportional to the star-formation rate (SFR).

\section*{GRBs as a Probe of Star Formation}

Observational estimates [16,17,18,19] indicate that
the SFR in the universe was about 15 times larger at a redshift $z
\approx 1$ than it is today.  The data at higher redshifts from the
Hubble Deep Field (HDF) in the north suggests a peak in the SFR at $z
\approx 1-2$ [19], but the actual
situation is highly uncertain.   However, theoretical calculations show
that the birth rate of Pop III stars produces a peak in the
SFR in the universe at redshifts $16 \lesssim z
\lesssim 20$, while the birth rate of Pop II stars produces a much
larger and broader peak at redshifts $2 \lesssim z \lesssim 10$
[20,21,22].  Therefore one expects GRBs to occur out to at least $z \approx
10$ and possibly $z \approx 15-20$, redshifts that are far larger than
those expected for the most distant quasars.  Consequently GRBs may be
a powerful probe of the star-formation history of the universe, and
particularly of the SFR at VHRs.  

\begin{figure}
\begin{minipage}[t]{2.75truein}
\mbox{}\\
\psfig{file=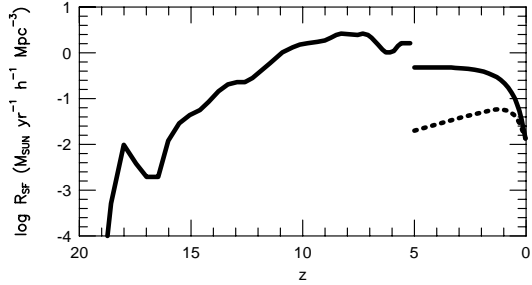,width=2.75truein,clip=}
\end{minipage}
\hfill
\begin{minipage}[t]{2.75truein}
\mbox{}\\
\caption{The cosmic SFR $R_{SF}$ as a function of
redshift $z$.  The solid curve at $z < 5$ is the SFR
derived by [23]; the solid curve at $z \ge 5$ is the 
SFR calculated by [21] (the dip in
this curve at $z \approx 6$ is an artifact of their numerical
simulation).  The dotted curve is the SFR derived by
[19].}
\end{minipage}
\end{figure}

In Figure 1, we have plotted the SFR versus redshift from a
phenomenological fit [23] to the SFR
derived from submillimeter, infrared, and UV data at redshifts $z < 5$,
and from a numerical simulation by [21] at redshifts  $z \geq 5$.  The simulations done by
[21] indicate that the SFR increases with
increasing redshift until $z \approx 10$, at which point it levels
off.  The smaller peak in the SFR at $z \approx 18$ corresponds to the
formation of Population III stars, brought on by cooling by molecular
hydrogen.  Since GRBs are detectable at these VHRs and their redshifts
may be measurable from the absorption-line systems and the Ly$\alpha$
break in the afterglows [24], if the GRB rate is
proportional to the SFR, then GRBs could provide unique
information about the star-formation history of the VHR universe.

More easily but less informatively, one can examine the GRB peak photon
flux distribution $N_{GRB}(P)$.  To illustrate this, we have calculated
the expected GRB peak flux distribution assuming (1) that the GRB rate
is proportional to the SFR\footnote{This may
underestimate the GRB rate at VHRs since it is generally thought that
the initial mass function will be tilted toward a greater fraction of
massive stars at VHRs because of less efficient cooling due to the
lower metallicity of the universe at these early times.}, (2) that the
SFR is that given in Figure 1, and (3) that the peak
photon luminosity distribution $f(L_P)$ of the bursts is independent of
$z$.  There is a mis-match of about a factor of three between the $z <
5$ and $z \geq 5$ regimes.  However, estimates of the star formation
rate are uncertain by at least this amount in both regimes.  We have
therefore chosen to match the two regimes smoothly to one another, in
order to avoid creating a discontinuity in the GRB peak flux
distribution that would be entirely an artifact of this mis-match.


For a peak luminosity function $f(L_P)$ and for $dL_P/d\nu \propto 
\nu^{-\alpha}$, the observed GRB peak flux distribution $N_{GRB}(P)$ is given by the following convolution integration:
\begin{equation}
N_{GRB}(P) = \Delta T_{obs} \int_0^{\infty}R_{GRB}(P|L_P)f[L_P-4\pi D^2(z)(1+z)^{\alpha}P]dL_P \; ,
\end{equation}
where $\Delta T_{obs}$ is the length of time of observation, $D(z)$ is comoving distance,
\begin{equation}
R_{GRB}(P|L_P) \propto  
\frac{R_{SF}(z)}{1+z}\frac{dV(z)}{dz}\left|\frac{dz(P|L_P)}{dP}\right|
\; ,
\end{equation}
$R_{SF}(z)$ is the local co-moving SFR at $z$, and $dV(z)/dz$ is  differential comoving volume [24].

\begin{figure}[t]
\psfig{file=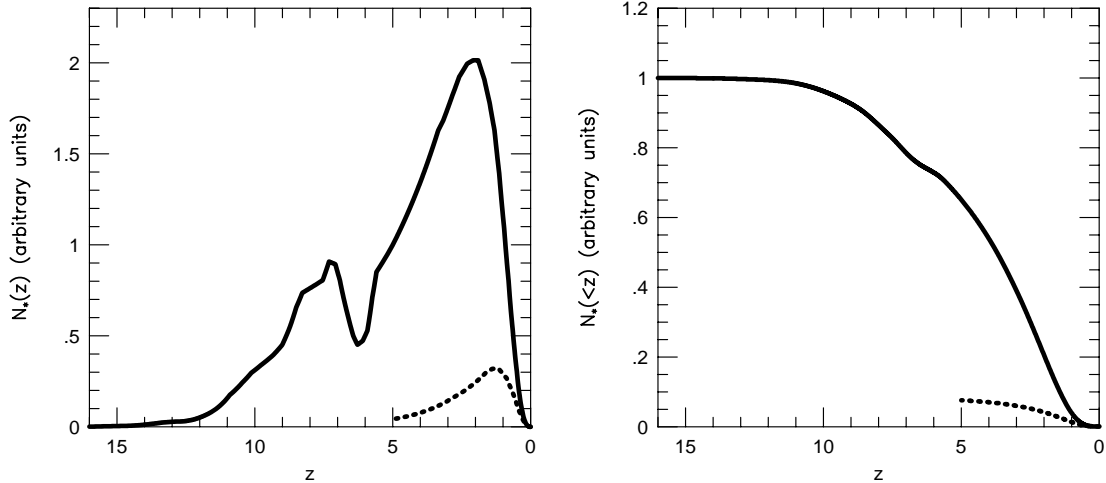,width=5.75truein,clip=} 
\caption{Top panel:  The number $N_*$ of stars expected as a function
of redshift $z$ (i.e., the SFR from Figure 1, weighted
by the differential comoving volume, and time-dilated) assuming that
$\Omega_M = 0.3$ and $\Omega_\Lambda = 0.7$.  Bottom panel:  The
cumulative distribution of the number $N_*$ of stars expected as a
function of redshift $z$.  Note that $\approx 40\%$ of all stars have
redshifts $z > 5$.  The solid and dashed curves in both panels have the
same meanings as in Figure 1.}
\end{figure}

The left panel of Figure 2 shows the number $N_*(z)$ of stars expected
as a function of redshift $z$ (i.e., the SFR, weighted
by the co-moving volume, and time-dilated) for an assumed cosmology
$\Omega_M = 0.3$ and $\Omega_\Lambda = 0.7$ (other cosmologies give
similar results).  The solid curve corresponds to the star-formation
rate in Figure 1.  The dashed curve corresponds to the star-formation
rate derived by [19].  This figure shows that $N_*(z)$
peaks sharply at $z \approx 2$ and then drops off fairly rapidly at
higher $z$, with a tail that extends out to $z \approx 12$.  The rapid
rise in $N_*(z)$ out to $z \approx 2$ is due to the rapidly increasing
volume of space.  The rapid decline beyond $z \approx 2$ is due almost
completely to the ``edge'' in the spatial distribution produced by the
cosmology.  In essence, the sharp peak in  $N_*(z)$ at $z \approx 2$
reflects the fact that the  SFR we have taken is fairly
broad in $z$, and consequently, the behavior of $N_*(z)$ is dominated
by the behavior of the co-moving volume $dV(z)/dz$; i.e., the shape of
$N_*(z)$ is due almost entirely to cosmology.  The right panel in
Figure 2 shows the cumulative distribution $N_*(>z)$ of the number of
stars expected as a function of redshift $z$.  The solid and dashed
curves have the same meaning as in the upper panel.  This figure shows
that $\approx 40\%$ of all stars have redshifts $z > 5$.

\begin{figure}
\begin{minipage}[t]{2.75truein}
\mbox{}\\
\psfig{file=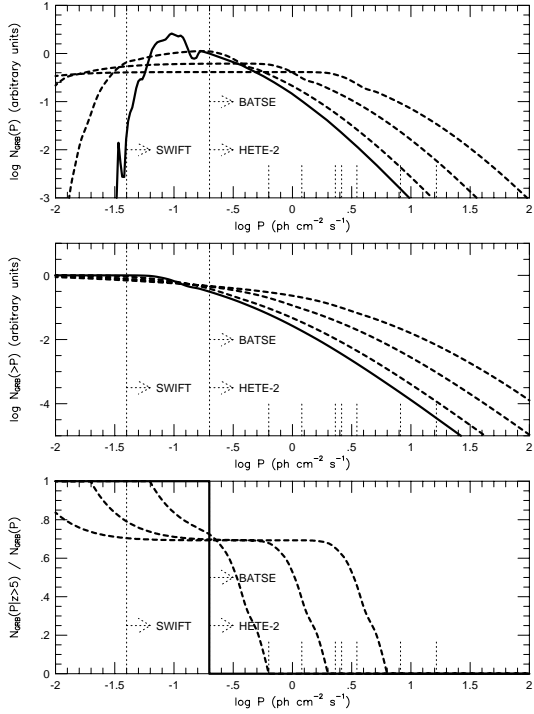,width=2.75truein,clip=}
\end{minipage}
\hfill
\begin{minipage}[t]{2.75truein}
\mbox{}\\
\caption{Top panel:  The differential peak photon flux distribution of
GRBs, assuming that (1) the GRB rate is proportional to the
SFR, (2) the SFR is that shown in
Figure 1; and (3) the bursts are standard candles with a peak photon
luminosity $L_P = 10^{58}$ ph cm$^{-2}$ s$^{-1}$ (solid curve), or have
a logarithmically flat peak photon luminosity function that spans a
factor of 10, 100, or 1000 (dashed curves).  Approximate detection
thresholds are plotted for BATSE and HETE-2, and for {\it Swift}
(dotted lines).  Middle panel:  The cumulative peak photon flux
distribution of GRBs for the same luminosity functions.  Lower panel: 
The fraction of GRBs with peak photon flux $P$ that have redshifts of
$z \ga 5$ for the same luminosity functions.  In all three panels, the 
dotted hashes mark the peak photon fluxes of the bursts with known peak
photon luminoisities and redshifts .}
\end{minipage}
\end{figure}

The upper panel of Figures 3 shows the predicted peak photon flux
distribution $N_{GRB}(P)$.  The solid curve assumes that all bursts
have a peak (isotropic) photon luminosity $L_P = 10^{58}$ ph s$^{-1}$. 
However, there is now overwhelming evidence that GRBs are not
``standard candles.''  Consequently, we also show in Figure 3, as an
illustrative example, the convolution of this same SFR
and a logarithmically flat photon luminosity function  $f(L_P)$
centered on $L_P = 10^{58}$ ph s$^{-1}$, and having widths $\Delta L_P
/ L_P = 10$, 100 and 1000.\footnote{The seven bursts with
well-determined redshifts and published peak (isotropic) photon
luminosities have a mean peak photon luminosity and sample variance
$\log L_P = 58.1 \pm 0.7$.}  The actual luminosity function of GRBs
could well be even wider [25]. 

The middle panel of Figure 3 shows the predicted cumulative peak photon
flux distribution $N_{GRB}(> P)$ for the same luminosity function.  For
the SFR that we have assumed, we find that, if GRBs are
assumed to be ``standard candles,'' the predicted peak photon flux
distribution falls steeply throughout the BATSE and HETE-2 regime, and
therefore fails to match the observed distribution, in agreement with
earlier work.  In fact, we find that a photon luminosity function
spanning at least a factor of 100 is required in order to obtain
semi-quantitative agreement with the principle features of the observed
distribution; i.e., a roll-over at a peak photon flux of $P \approx 6$
ph cm$^{-2}$ s$^{-1}$ and a slope above this of about -3/2.  This
implies that there are large numbers of GRBs with peak photon number
fluxes below the detection threshold of BATSE and HETE-2, and even of
{\it Swift}.  

The lower panel of Figure 3 shows the predicted fraction of bursts with
peak photon number flux $P$ that have redshifts of $z > 5$, for the
same luminosity functions.  This panel shows that a significant fraction
of the bursts near the {\it Swift} detection threshold will have
redshifts of $z > 5$.

\section*{Conclusions}

We have shown that, if many GRBs are indeed produced by the collapse of
massive stars, one expects GRBs to occur out to at least $z \approx 10$
and possibly $z \approx 15-20$, redshifts that are far larger than
those expected for the most distant quasars.  GRBs therefore give us
information about the star-formation history of the universe, including
the earliest generations of stars.  The absorption-line systems and the
Ly$\alpha$ forest visible in the spectra of GRB afterglows can be used
to trace the evolution of metallicity in the universe, and to probe the
large-scale structure of the universe at very high redshifts.  Finally,
measurement of the Ly$\alpha$ break in the spectra of GRB afterglows
can be used to constrain, or possibly measure, the epoch at which
re-ionization of the universe occurred, using the Gunn-Peterson test. 
Thus GRBs and their afterglows may be a powerful probe of the very high
redshift ($z \gtrsim 5$) universe.

\end{document}